%% file: main.tex
\RequirePackage{lineno}
\documentclass[twocolumn,showpacs,superscriptaddress,amsmath,amssymb,nofootinbib]{revtex4-2}

\usepackage{graphicx}
\usepackage{subfigure}
\usepackage{subcaption}
\usepackage{epsfig}
\usepackage{overpic}
\usepackage{dcolumn}
\usepackage[normalem]{ulem}
\usepackage{bm}
\usepackage{color}
\usepackage{lineno}
\usepackage{xspace}
\usepackage{multirow}
\usepackage{epstopdf}
\usepackage{xcolor}
\usepackage{soul}
\usepackage{verbatim}
\usepackage{enumitem}
\usepackage{todonotes}
\usepackage{cancel}
\usepackage{array}
\usepackage[toc,page]{appendix}
\usepackage[colorlinks,allcolors=black]{hyperref}
\usepackage{multirow}
\usepackage{ytableau}
\usepackage{comment}

\usepackage{diagbox}
\usepackage{titlesec}
\titleformat{\section}{\centering\normalfont\normalsize\bfseries}{\thesection}{1em}{}

\newcommand{\sfim}{\affiliation{Guangdong Basic Research Center of Excellence for Structure and Fundamental Interactions of Matter, Guangdong Provincial Key Laboratory of Nuclear Science, Guangzhou 510006, China}}

\newcommand{\ihep}{\affiliation{Institute of High Energy Physics, Chinese Academy of Sciences, Beijing 100049, China}}

\newcommand{\ccast}{\affiliation{China Center of Advanced Science and Technology, Beijing 100190, China}}

\newcommand{\iqm}{\affiliation{State Key Laboratory of Nuclear Physics and Technology, Institute of Quantum Matter, South China Normal University, Guangzhou 510006, China}}

\newcommand{\scnt}{\affiliation{Southern Center for Nuclear-Science Theory (SCNT), Institute of Modern Physics, Chinese Academy of Sciences, Huizhou 516000, Guangdong Province, China}}

\newcommand{\gscas}{\affiliation{Graduate School of China Academy of Engineering Physics, Beijing 100193, China}}

\newcommand{\ucas}{\affiliation{University of Chinese Academy of Sciences, Beijing 100049,   China}}

\begin{document}

\title{\boldmath Fully charmed tetraquarks on the Lattice with controlled errors}

\author {Zhenyu Zhang}
\email{zhyzhang@ihep.ac.cn}
\ihep
\ccast
\author {Bing-Nan Lu}
\email{bnlv@gscaep.ac.cn}
\gscas
\author {Qian Wang}
\email{qianwang@m.scnu.edu.cn}
\iqm
\sfim
\scnt
\author {Qiang Zhao}
\email{zhaoq@ihep.ac.cn}
\ihep\ucas

\date{\today}

\begin{abstract}
We introduce the lattice quark potential model (LQPM), which discretizes the nonrelativistic quark potential model on a periodic cubic lattice and diagonalizes the resulting many-body Hamiltonian exactly. Unlike variational approaches, which provide only upper bounds on the energy, LQPM removes this variational bias, and its remaining finite-volume uncertainty is removed by extrapolation. Formulated directly in coordinate space, the method is indifferent to the specific form of the potential and accommodates arbitrary interactions, such as including three-body forces and spin-orbit and tensor couplings, that are difficult to treat in conventional few-body approaches. As a benchmark, the charmonium and $\Omega_{ccc}$ spectra are reproduced within a few MeV of experiment. For the fully charmed tetraquark $cc\bar{c}\bar{c}$, we obtain three resonances $J^P = 0^+, 1^+,$ and $2^+$ ground states above the $\eta_c \eta_c$, $\eta_c J/\psi$, $J/\psi J/\psi$ thresholds, respectively.
%we obtain a $J^P=0^+$ ground state at $5934.8(13)~\mathrm{MeV}$, $33.2(13)~\mathrm{MeV}$ below the $\eta_c\eta_c$ threshold, unambiguously identified as a deeply bound tetraquark whose existence is robust against the choice of potential. Such a state can decay into two dileption pairs $2(e^+e^-)$ via $J/\psi e^+e^-$, with significance similar to $X(6400)$, or about one order of magnitude smaller. Lying below the lowest strong-decay channel, it can decay only through electromagnetic and weak interactions. 
Our results establish exact lattice diagonalization as a controlled, systematically improvable, and broadly applicable route to multiquark spectroscopy, providing a quantitative benchmark for the strong interaction and valuable guidance for future experimental searches for exotic hadrons.
\end{abstract}

\maketitle

% keywords can be removed
{\it Introduction}~~
Color confinement in quantum chromodynamics (QCD) dictates that only
color-neutral objects, i.e. hadrons, can be the most direct observables in experiment. Hadron spectroscopy therefore provides a direct window into the dynamics of the strong interaction. In the conventional quark model~\cite{Gell-Mann:1964ewy,Zweig:1964ruk}, mesons are built from
$q\bar q$ pairs and baryons from three quarks $qqq$. Confinement, however, admits a
far richer spectrum: color-singlet multiquark configurations, known as exotic
hadrons, are equally allowed. Lying outside the minimal valence-quark picture,
these states provide a unique testing ground for QCD.
Since the observation of the $X(3872)$~\cite{Belle:2003nnu}, dozens of exotic
hadrons have been discovered, including the doubly charmed
tetraquark $T_{cc}^+$~\cite{LHCb:2021auc,LHCb:2021vvq}, the charged
charmonium-like state $Z_c(3900)$~\cite{BESIII:2013ris,Belle:2013yex}, the
hidden-charm pentaquarks $P_c$~\cite{LHCb:2015yax,LHCb:2016ztz,LHCb:2019kea},
the strange pentaquark $P_{cs}$~\cite{LHCb:2022ogu}, and the fully charmed
tetraquark candidate $X(6900)$~\cite{LHCb:2020bwg} and so on. Their very existence has
reshaped our picture of the hadronic spectrum, yet their internal
structure---ranging from loosely bound hadronic molecules to compact multiquark
states---remains under intense debate~\cite{Guo:2017jvc,Liu:2019zoy,Chen:2016qju,Esposito:2016noz,Brambilla:2019esw,Chen:2022asf,Guo:2019twa,Richard:2016eis,Ali:2017jda}.
Resolving this question is one of the central challenges of modern hadron
physics.

Fully-heavy multiquark systems, containing only charm and bottom quarks, are
particularly clean laboratories for such studies~\cite{Liu:2019zuc,Liang:2024met}.
With no light quarks present, the long-range pion exchange is absent. Thus, any binding
must arise from short-range QCD
forces~\cite{Brambilla:2019esw,Esposito:2016noz,Liu:2019zuc,Liang:2024met},
which favors compact multiquark states over hadronic molecules. Their spectra
can therefore be obtained directly from nonrelativistic quark models in which
the dynamics is governed by one-gluon-exchange (OGE) and confinement
potentials~\cite{Liu:2019zuc,Liang:2024met,Wang:2019rdo,Yan:2021glh,Hiyama:2018ukv},
i.e., by solving the nonrelativistic many-body Schr\"odinger equation. The constituent quark model is expected to be reliable here, for the charm quark $\hbar c/m_c\approx0.13$ fm is much smaller than the hadronic size, and the leading order terms in the $1/m_Q$ expansion of QCD are a nonrelativistic kinetic energy together with a static two-body potential~\cite{Brambilla:2004wf}, whose short- and long-range parts follow from OGE and the static quark--antiquark potential measured in lattice QCD~\cite{Bali:2000gf,Lucha:1991vn}, respectively. The neglected gluonic, meson--meson and higher-order corrections are less important for fully-heavy bound states. The difficulty therefore lies not in the Hamiltonian but in solving it. Variational basis expansions, such as the Gaussian
expansion method (GEM)~\cite{Hiyama:2003cu} and the explicitly correlated
Gaussian (ECG) method~\cite{Varga:1995dm,Mitroy:2013eom}, suffer from the
exponential growth of the basis size with the number of particles and provide
only an upper bound on the energy. While diffusion Monte Carlo (DMC) approaches
face the fermion sign problem~\cite{Gordillo:2020sgc,Ma:2023int,Gordillo:2024blx}.
More recently, the neural-network quantum state (NNQS)
approach~\cite{Wu:2025wvv,Li:2026hol} has emerged as a promising tool for
multiquark states, and we have applied it to fully-heavy
multiquarks~\cite{Li:2026hol}; DeepQuark~\cite{Wu:2025wvv} can circumvent the
computational barriers of conventional methods. The advantage of NNQS lies in
its compact parametrization of the many-body wave function, which avoids the
exponential growth of a fixed basis. Yet it remains variational in nature and
therefore yields only an upper bound on the energy, leaving the distance to the
true solution uncontrolled.

Hadronic lattice effective field theory (HLEFT)~\cite{Zhang:2024yfj} is an
extension of nuclear lattice effective field theory (NLEFT)~\cite{Lahde:2019npb,Lee:2025req,Lee:2008fa},
an established powerful framework for nuclear \textit{ab initio} calculations that
combines chiral effective field theory with lattice quantum Monte Carlo methods or direct diagonalization to tackle the intricate nuclear many-body problems. NLEFT has been successfully applied
for exploring nuclear structure and interactions, including ground
states~\cite{Borasoy:2006qn,Epelbaum:2009pd,Epelbaum:2010xt,Lahde:2013uqa,Lu:2018bat,Lu:2021tab,Elhatisari:2022zrb}
and excited states~\cite{Epelbaum:2013paa,Meissner:2023cvo}, intrinsic
density distributions and clustering
phenomena~\cite{Lahde:2013png,Epelbaum:2011md,Epelbaum:2012qn,Epelbaum:2012iu,Elhatisari:2017eno,Zhang:2024wfd, Shen:2022bak,Shen:2024qzi},
nucleus-nucleus
scattering~\cite{Bour:2012hn,Elhatisari:2015iga}, nuclear
matter at zero and finite
temperature~\cite{Elhatisari:2016owd,Lu:2019nbg,Ren:2023ued,Ma:2023ahg} and, more recently,
hypernuclei~\cite{Hildenbrand:2024ypw,Haidenbauer:2025zrr,Tong:2025sui,Tong:2024jvs}. Built on the same lattice machinery, HLEFT carries this proven power to the hadronic few- and many-body problem, opening a promising \textit{ab initio} route to multiquark spectra starting from the QCD-based EFT Hamiltonian.

In this work we formulate the lattice quark potential model (LQPM), which is the
nonrelativistic quark potential model discretized on a periodic cubic lattice,
and the resulting Hamiltonian is diagonalized exactly. This yields the exact
eigenvalues of the discretized Hamiltonian (up to numerical precision) rather
than a variational upper bound, which is the essential difference from the GEM
and NNQS approaches. The residual uncertainties, dominated by the finite box
size, are controlled by finite-volume extrapolation. We then apply LQPM to fully
charmed multiquark systems. Restricting the flavor content to charm has three
advantages. One is that long-range pion exchange is absent, so that the dynamics is
dominated by the OGE and confinement potentials. Secondly the parameters of these
potentials are already fixed by the charmonium spectrum, so that predictions for
multiquark systems introduce no new parameters. At the end, relativistic and three-body
corrections are suppressed by powers of the heavy-quark mass~\cite{Wu:2025wvv}. As a first
application, we benchmark our results against NNQS and GEM for the charmonia and
$\Omega_{ccc}$. We then study the fully charmed tetraquark, for which we obtain three resonances $J^P = 0^+, 1^+,$ and $2^+$ ground states above the $\eta_c \eta_c$, $\eta_c J/\psi$, $J/\psi J/\psi$ thresholds, respectively.

\begin{figure*}[htbp]
    \centering
    \includegraphics[width=0.32\textwidth]{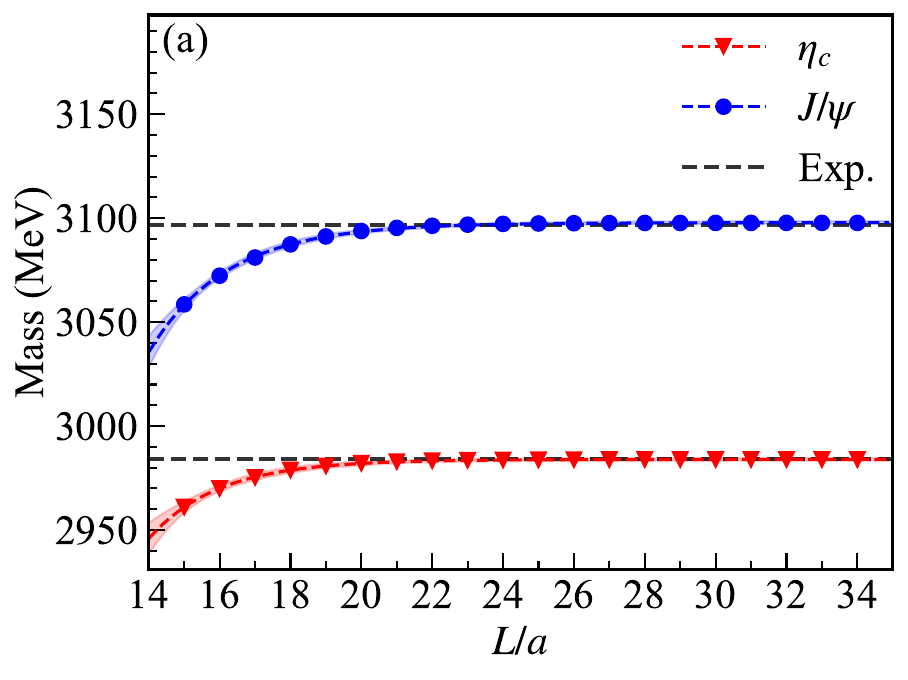}
    \includegraphics[width=0.32\textwidth]{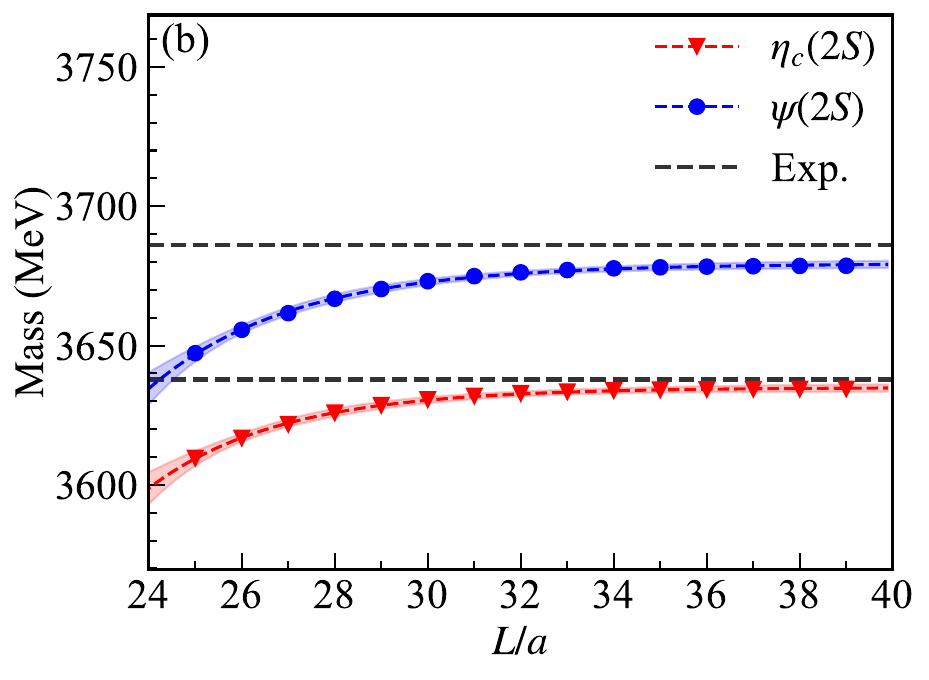}
    \includegraphics[width=0.32\textwidth]{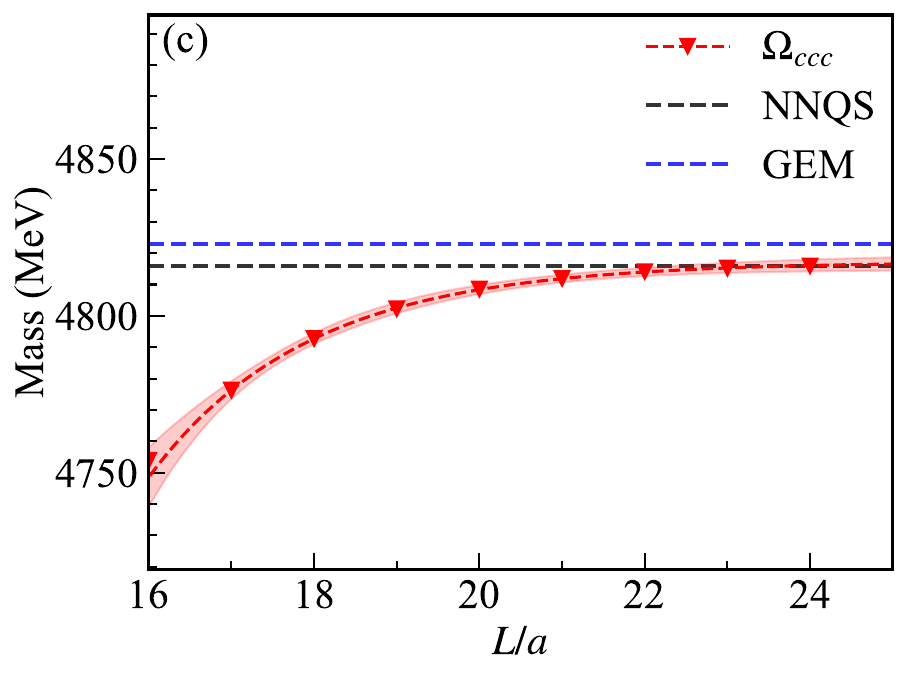}
    \caption{A comparison of the predicted masses of the charmonium ground states (a), first excited states (b), and $\Omega_{ccc}$ (c). The red triangles and blue points represent this work predictions. The finite-volume errors are shown by the shaded area. Black dashed lines in subfigure (a) and (b) represent the experimental data~\cite{ParticleDataGroup:2020ssz}. Black dashed line and blue dashed line in subfigure (c) represent the NNQS results~\cite{Li:2026hol} and the GEM prediction by Zhong et al., in Ref.~\cite{Liu:2019zuc}, respectively.
    }
    \label{fig:charmonium}
\end{figure*}

{\it Method}~~
Various potentials have been used to study hadron
spectroscopy~\cite{Lucha:1991vn,Eichten:1978tg,Brambilla:2004wf,Bali:2000gf},
and in particular multiquark
spectroscopy~\cite{Liu:2019zuc,Liang:2024met,Wang:2019rdo,Yan:2021glh,Hiyama:2018ukv,Karliner:2022xtd,Olsen:2017bmm,Karliner:2016zzc,Zouzou:1986}.
We describe fully charmed multiquark systems with the nonrelativistic quark
potential model (NRQPM),
\begin{equation}
    H = \sum_{i}^{n}(m_i+ T_i ) + \sum_{i<j}^{n}V_{ij}(r_{ij}),
\end{equation}
where $m_i$ and $T_i$ denote the mass and kinetic energy of the $i$-th
(anti)quark, and $V_{ij}(r_{ij})$ is the effective potential between
(anti)quarks $i$ and $j$, depending on their relative distance
$r_{ij}=|\bm r_i-\bm r_j|$. The potential contains an OGE term $V_{ij}^{\text{OGE}}$ and a linear
confinement term $V_{ij}^{\text{Conf}}$~\cite{Liu:2019zuc,Liang:2024met},
\begin{align}
    V_{ij}^{\text{OGE}} =& \frac{\alpha_{ij}}{4}(\bm{\lambda}_i \cdot \bm{\lambda}_j)\left [ \frac{1}{r_{ij}}-\frac{\pi}{2} \frac{\sigma_{ij}^{3}e^{-\sigma_{ij}^{2}r_{ij}^2}}{\pi^{3/2}}\frac{4}{3m_i m_j}(\bm{\sigma}_i\cdot\bm{\sigma}_j) \right ], \nonumber\\
    V_{ij}^{\text{Conf}}=&-\frac{3}{16}(\bm{\lambda}_i \cdot \bm{\lambda}_j)  br_{ij},
    \label{eq:potential}
\end{align}
where $\bm{\lambda}_i$ is the Gell-Mann matrix acting on the $i$-th quark
(replaced by $-\bm{\lambda}^*$ for antiquarks) and $\bm{\sigma}_i$ the Pauli
matrix. The parameters $\alpha_{ij}$ and $\sigma_{ij}$ are the strong coupling
and the Gaussian regulator of the spin-spin interaction for the flavor pair
$(i,j)$, and $b$ is the strength of the linear
confinement.\footnote{A cutoff is introduced to resolve the singularity of the
$1/r_{ij}$ term in Eq.~\eqref{eq:potential}, similar to that in
Ref.~\cite{Niu:2025uxk}.} The parameters are determined by fitting to the charmonium
spectrum~\cite{Liu:2019zuc,Liang:2024met} in continuous space and are collected
in Table~\ref{tab:parameters}. For all-charmed hadrons, three-body interaction
effects are suppressed by the heavy-quark mass and are therefore
neglected~\cite{Wu:2025wvv}; nevertheless, our method can incorporate
three-body or four-body forces straightforwardly if necessary. In the absence of
spin-orbit and tensor operators, both the total orbital angular momentum $L$ and
the total spin $S$ are good quantum numbers. 

\begin{table}[htbp]
\setlength{\tabcolsep}{3.5mm}
\renewcommand{\arraystretch}{1.5}
\centering
\caption{Parameters of the model used~\cite{Liu:2019zuc,Liang:2024met}.}
\label{tab:parameters}
\begin{tabular}{cccc}
\hline
\hline
$m_c$ (GeV) & $\alpha_{cc}$ & $\sigma_{cc}$ (GeV) & $b$ (GeV$^2$)\\
\hline
 1.483      & 0.5461        & 1.1384              & 0.1425\\
\hline
\hline
\end{tabular}
\end{table}

The total wave function of multiquark can be expressed as a direct product of the flavor, spatial, spin, and color wave functions:
\begin{align}
    \Psi_{\text{total}} = \Psi_{\text{spatial}}\otimes \Psi_{\text{color}}\otimes\Psi_{\text{spin}}\otimes\Psi_{\text{flavor}}.
\end{align}
In this work, the color and spin wave functions are constructed via $\text{SU(3)} \otimes \text{SU(2)}$ symmetry, and the complete spatial wave function is defined on lattice coordinate space. Additionally, the flavor wave function of fully charmed multiquark is symmetric. Similar to Ref.~\cite{Wu:2025wvv}, Fermi–Dirac statistics for identical (anti)quarks is implemented through the exchange symmetry of the spatial wave function: for every admissible color–spin configuration, the spatial part is assigned the exchange symmetry that renders the total wave function antisymmetric under the interchange of any two identical (anti)quarks.

 The color-spin configurations of charmonia and of the fully charmed baryon
$\Omega_{ccc}$ are well established. For fully charmed tetraquarks, we adopt the
$(cc)(\bar{c}\bar{c})$ configuration~\footnote{Note that this is not a
diquark-antidiquark approximation, but rather a way of combining the quarks into
a color-neutral object.}, for which the color-spin degrees of freedom can be
coupled into the following independent bases:
\begin{align}
    \chi_{6\otimes\bar{6}}\phi^{s,s'}= & \left|(cc)_6^s(\bar{c}\bar{c})_{\bar{6}}^{s'}\right\rangle_1^S,\\
    \chi_{\bar{3}\otimes3}\phi^{s,s'}= & \left|(cc)_{\bar{3}}^s(\bar{c}\bar{c})_3^{s'}\right\rangle_1^S.
    \label{eq:singlet}
\end{align}
where $s$ and $s'$ run over all possible spin combinations. Here $\chi$ and
$\phi$ denote the color and spin bases, respectively, and the
corresponding matrix elements for all combinations are given in
Ref.~\cite{Liu:2019zuc}.

\begin{figure*}[htbp]
    \centering
    \includegraphics[width=0.32\textwidth]{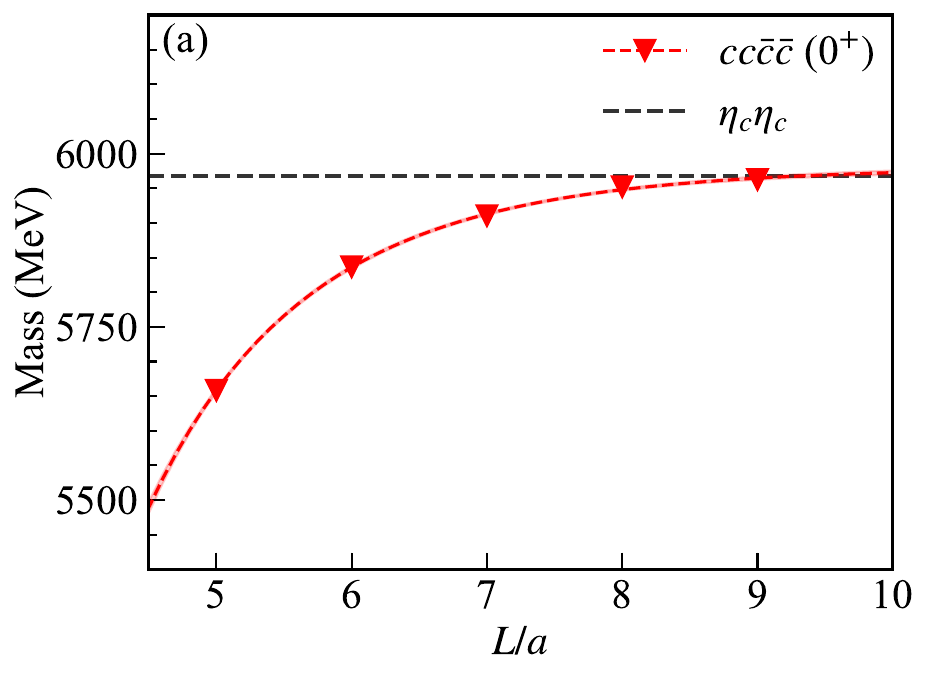}
    \includegraphics[width=0.32\textwidth]{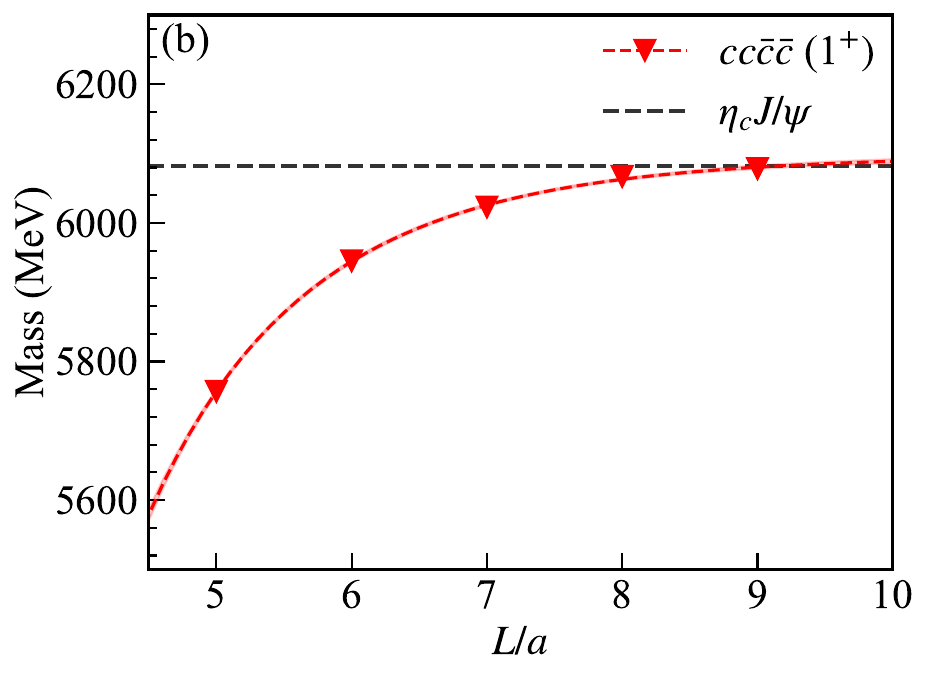}
    \includegraphics[width=0.32\textwidth]{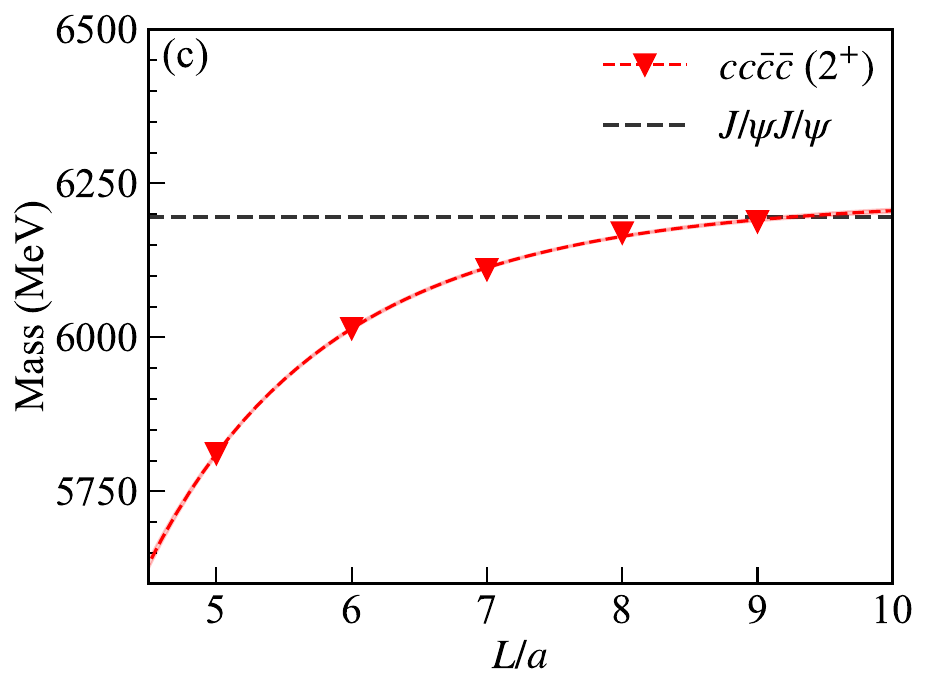}
    \caption{The predicted masses of S-wave fully charmed tetraquark  ground states with $J^P=0^+$(a), $1^+$ (b), and $2^+$ (c). The red triangles represent this work predictions with lattice spacing $a=1/1000$ MeV$^{-1}$. The finite-volume errors are shown by the shaded area. The black dashed lines in subfigure (a), (b) and (c) represent the $\eta_c\eta_c$, $\eta_cJ/\psi$, and $J/\psi J/\psi$ thresholds calculated in this framework, respectively.}
    \label{fig:tetraquark}
\end{figure*}

For the spatial part, we discretize the Hamiltonian on a cubic box with $L^3$ sites 
and lattice spacing $a$, imposing periodic boundary conditions. The single-particle momenta are $\bm p=2\pi\bm n/L$ where $\bm n$ is integer vector. For an $N$-body system we retain only the $N-1$
relative coordinates and remove the center-of-mass motion by enforcing vanishing total momentum, $\sum_i^N \bm p_i=0$. The resulting Hilbert space has dimension $L^{3(N-1)}$. 
Since the two-body potentials depend only on the distance between the two particles, they are local in coordinate space and can be applied directly on the lattice. The kinetic energy is evaluated in momentum space: after a fast Fourier transform, the single-particle momenta are applied directly to the wave function, which is then transformed back to coordinate space. The Schrödinger equation is therefore solved by direct diagonalization of the Hamiltonian, using the implicitly restarted Lanczos method~\cite{lehoucq:1998} as implemented in SciPy~\cite{Virtanen:2019joe} in Python.
Because the spatial wave function is defined as a full wave function in lattice coordinate space, it contains contributions from all orbital angular momenta; after exact diagonalization, the ground-state wave function is naturally an S wave.
\begin{table}[htbp]
\setlength{\tabcolsep}{2mm}
\renewcommand{\arraystretch}{1.5}
\centering
\caption{Masses (MeV) of the charmonia and all-charmed baryons. Our results are compared with NNQS results~\cite{Li:2026hol}, GEM results by Zhong et al.~\cite{Liu:2019zuc,Liang:2024met} and experimental data~\cite{ParticleDataGroup:2020ssz}. The errors of this work come from fitting with the finite volume formula.}
\begin{tabular}{lccccc}
\hline
\hline
State & $J^P$ & This work & NNQS & GEM & Exp.\\
\hline
$\eta_c$     & $0^-$    & 2984.0(2) & 2983.7(10) & 2983 & 2984\\
$J/\psi$     & $1^-$    & 3097.8(2) & 3097.2(8)  & 3097 & 3097\\
$\eta_c(2S)$ & $0^-$    & 3634.9(5) & $\cdots$   & 3635 & 3638\\
$\psi(2S)$   & $1^-$    & 3679.4(5) & $\cdots$   & 3679 & 3686\\
$\Omega_{ccc}$ & $3/2^+$ & 4817.3(9) & 4816.1(5)  & 4823 & $\cdots$\\
\hline
\hline
\end{tabular}
\label{tab:charmonium}
\end{table}

{\it Results}~~
The lattice method is first benchmarked against the charmonia and $\Omega_{ccc}$, for which the charmonium masses are known precisely. The calculations are performed at a lattice spacing 
$a=1/2000$ MeV$^{-1}\approx0.1$ fm with the meson box sizes ranging from $L^3=1.5^3$ to 3.9$^3$ fm$^3$, and the baryon box sizes from $L^3=1.5^3$ to 2.4$^3$ fm$^3$. For each state, the results are extrapolated to the physical point using the finite-volume extrapolation formula~\cite{Luscher:1985dn,Zhang:2025juc}; different finite-volume formulas yield consistent results. The quantum numbers of $\eta_c(2S)$ and $\psi(2S)$ are obtained through standard angular-momentum projection~\cite{Zhang:2024yfj,Lu:2014xfa}; see the supplementary materials for further details. Figure~\ref{fig:charmonium} and Table~\ref{tab:charmonium} show the lattice results together with the finite-volume fits and their uncertainties, and compare them with the results from NNQS, GEM, and experiment.

For the charmonium ground states, i.e. $\eta_c$ and $J/\psi$, their masses agree with
experiment to within a few MeV and are consistent with the
NNQS~\cite{Li:2026hol} and GEM~\cite{Liu:2019zuc,Liang:2024met} results obtained
with the same Hamiltonian. The first radial excitations, $\eta_c(2S)$ and
$\psi(2S)$, converge only at larger box sizes than the ground states. Their
masses differ from the experimental values by a few MeV but remain consistent
with the GEM results. This agreement validates our lattice implementation of the
kinetic, confinement, and OGE terms of the Hamiltonian.
The $\Omega_{ccc}$ mass converges to $4817.3(9)$ MeV, in agreement with the
NNQS result and below the GEM result. Since the GEM value is a variational upper
bound, its being higher than the converged lattice value is exactly what one
expects. It also confirms that our result sits very close to the exact eigenvalue.
The origin of this improvement is clear: lattice diagonalization solves the
discretized Hamiltonian exactly and is not constrained by the functional form of
a trial wave function, whereas variational methods can only return an upper
bound on the energy. Moreover, the errors in LQPM are explicit and can be
systematically reduced, so its accuracy is under quantitative control rather
than merely bounded from one side.

\begin{table*}[htbp]
\setlength{\tabcolsep}{6mm}
\renewcommand{\arraystretch}{1.5}
\centering
\caption{Predicted masses (MeV) of fully charmed tetraquark $cc\bar{c}\bar{c}$ in lattice spacing $a=1/1000$ MeV$^{-1}$ and compare them with the corresponding thresholds (MeV) computed in the
same framework. The error of this work comes from the finite volume effect.}
\label{tab:tetraquark}
\begin{tabular}{ccccccc}   
\hline
\hline
{State}&{$J^{P}$} & {Thresholds} &  $L/a$ & {Mass} &$\chi_{6\otimes\bar{6}}$&$\chi_{\bar{3}\otimes3}$\\
\hline
\multirow{6}{*}{$cc\bar{c}\bar{c}$}
&\multirow{2}{*}{$0^{+}$}  & \multirow{2}{*}{$\eta_c\eta_c=5968$} &9& 5962.5& $\sim 65.1\%$& $\sim 34.9\%$\\
&&&$\infty$& 5980.7(14) & $\cdots$&$\cdots$\\
&\multirow{2}{*}{$1^{+}$}  & \multirow{2}{*}{$\eta_cJ/\psi=6082$} &9& 6078.4& $\sim 61.6\%$& $\sim 38.4\%$\\
&&&$\infty$& 6097.6(14) & $\cdots$&$\cdots$\\
&\multirow{2}{*}{$2^{+}$}  & \multirow{2}{*}{$J/\psi J/\psi=6196$}&9& 6187.9& $\sim 58.7\%$& $\sim 41.3\%$\\
&&&$\infty$& 6225.5(19) & $\cdots$&$\cdots$\\
\hline
\hline
\end{tabular}
\end{table*}

We next apply the method to fully charmed tetraquarks $cc\bar c\bar c$.
Balancing computational cost against accuracy, we use a lattice spacing
$a=1/1000$ MeV$^{-1}\approx0.2$ fm with box sizes $L^3=1.0^3$--$1.8^3$ fm$^3$, a range
sufficient for converged results. To reduce the dependence of results on the lattice spacing, we refitted the parameters in Table~\ref{tab:parameters} to charmonium spectrum when the lattice spacing $a=1/1000$ MeV$^{-1}$. The infinite-volume limit is reached by
extrapolating with the finite-volume
formula~\cite{Meissner:2014dea,Zhang:2026bqa}. Figure~\ref{fig:tetraquark} and
Table~\ref{tab:tetraquark} display the lattice results for each spin, together
with the finite-volume extrapolation, the  uncertainty, and the relevant thresholds.
The lightest $J^P=0^+$ state is predicted
at $5980.7(14)$ MeV, at least $12.7(14)$ MeV above the $\eta_c\eta_c$ threshold computed in the same framework. The calculation for this resonance is consistent with previous model calculations~\cite{Richard:2017vry,Liu:2019zuc,Li:2026hol,Faustov:2021hjs,Yu:2022lak,Yan:2023lvm,Wu:2016vtq,Chen:2016jxd,Gordillo:2020sgc,Yang:2021hrb,Lu:2020cns,Zhang:2022qtp,Wang:2019rdo,Wu:2025wvv,Barnea:2006sd,Deng:2020iqw}. It might easily decay into $\eta_c \eta_c$ through the quark rearrangements.
%, and is thus identified as a fully charmed tetraquark bound state. Because the state is deeply bound, this conclusion is robust against the choice of potential and stands in sharp contrast to previous model calculations~\cite{Liu:2019zuc,Li:2026hol,Faustov:2021hjs,Yu:2022lak,Yan:2023lvm,Wu:2016vtq,Chen:2016jxd,Gordillo:2020sgc,Yang:2021hrb,Lu:2020cns,Zhang:2022qtp,Wang:2019rdo,Wu:2025wvv,Deng:2020iqw}, making the existence of a $0^{+}$ bound state below the $\eta_c\eta_c$ threshold a distinguishing feature of our framework. 
%Among all the allowed decay channels, $J/\psi \gamma$ dominates. Considering the capability of experimental detection, a four-lepton final state, in which one lepton pair originates from the $J/\psi$ and the others are from a virtual photon, may be easier to access. This region is complementary to that probed by the existing di-$J/\psi$ searches at CMS and ATLAS~\cite{LHCb:2020bwg,ATLAS:2023bft,CMS:2025fpt}. Furthermore, there are also other decay modes, such as  $\omega J/\psi$, $\phi J/\psi$, $\eta \eta_c$, together with the semileptonic modes $e^+ \nu_eJ/\psi D_s^-$ and its charge conjugate channel $e^- \bar{\nu}_eJ/\psi D_s^+$. 
%{\color{red}\it ZQ: The possible decay channel $\gamma J/\psi$ is difficult for LHCb, but easier for CMS in $J/\psi e^+e^-$. All the other listed channels should be much smaller. We should provide an estimate of the detection significance at CMS. }

The $J^P=1^+$ state at $6097.6(14)$ MeV lies only $15.6(14)$ MeV
above the $\eta_c J/\psi$ threshold. It might easily decay into $\eta_c J/\psi$ through the quark rearrangements.
%Since the binding energy is so small, this assignment may depend on the choice of potential, and its existence as a bound state is therefore less firm. Whether the state is a bound state or a resonance, it can be investigated in the $\eta_c e^+e^-$ channel. 
At the same time,  the $J^P=2^+$ state at
$6225.5(19)$ MeV is $29.5(19)$ MeV above the $J/\psi J/\psi$ threshold, and appears as a resonance. The interpretation is, however, unambiguous: our diagonalization approaches the true eigenvalue from below, identifying the state as a resonance. 
The state has already expressed itself as a near-threshold peak with quantum numbers $J^P = 2^+$ in the di-$J/\psi$ invariant mass distribution in the CMS analysis~\cite{CMS:2025fpt}. This structure is also interpreted as a near-threshold pole, denoted $X(6200)$, in coupled-channel analyses~\cite{Dong:2020nwy,Gong:2020bmg,Song:2024ykq,Liang:2021fzr}. On the other hand, the treatment of resonances remains a weak point of the present framework, as the resonance wave function cannot be normalized, making large finite-volume effect. A better determination would require, for instance, the complex scaling method with coupled channels~\cite{Wang:2022yes,Wang:2023jqs}.
Taken together, these results imply that the observed $X(6900)$ cannot be an 
S-wave ground-state fully charmed tetraquark.
Furthermore, the color-configuration ratio $\chi_{6\otimes\bar{6}}:\chi_{\bar{3}\otimes3}$
(Table~\ref{tab:tetraquark}) is strongly affected by finite-volume effects. In
the infinite-volume limit it converges to $2:1$ in accordance with the meson-meson scattering states~\cite{Wu:2024euj}, consistent with the DeepQuark
result~\cite{Wu:2025wvv}. Further details are given in the Supplementary
Material.

We compare our results with other model calculations. Our predicted masses are
significantly lower than those of the nonrelativistic quark models of
Refs.~\cite{Liu:2019zuc,Li:2026hol}, which employ the same Hamiltonian. The
origin of the discrepancy is clear. Those models construct color-spin wave
functions that are antisymmetric only approximately, together with symmetric
spatial wave functions, whereas we enforce full antisymmetry by explicitly
summing over all permutation combinations. As a result, higher partial waves also contribute to the ground-state energy (see the
Supplementary Material for details). Our results are likewise lower than the
relativistic quark model~\cite{Faustov:2021hjs,Yu:2022lak}, the MIT bag
model~\cite{Yan:2023lvm}, the color-magnetic diquark
model~\cite{Wu:2016vtq}, moment QCD sum rules~\cite{Chen:2016jxd},
DMC~\cite{Gordillo:2020sgc}, lattice-QCD insipired quark model~\cite{Yang:2021hrb}, the extended relativized quark
model~\cite{Lu:2020cns}, and several other potential-model
calculations~\cite{Zhang:2022qtp,Wang:2019rdo}.
In general, our results approach those of
DeepQuark~\cite{Wu:2025wvv}, hyperspherical formalism~\cite{Barnea:2006sd}, the color-magnetic interaction
model~\cite{Deng:2020iqw}, QCD sum rules~\cite{Wang:2018poa}, and the
nonrelativistic model~\cite{Debastiani:2017msn} from below. It is worth noting
that DeepQuark~\cite{Wu:2025wvv} uses a more general wave function that enforces
overall antisymmetry, and its fully charmed tetraquark states lie close to the
di-meson threshold. Because it is also a variational method, however, DeepQuark yields only an
upper bound on the energy, converging to the true value from above. Our exact diagonalization instead returns the eigenvalue of the discretized Hamiltonian directly.

{\it Summary and outlook}~~
In this work we have developed the LQPM and applied it, for the first time, to fully charmed hadrons. The nonrelativistic quark potential model is solved by discretizing the Schr\"odinger equation on a three-dimensional spatial lattice, and the Hamiltonian is diagonalized directly with the implicitly restarted Lanczos method; the systematic error originates from the finite-volume effect and is under control. We have benchmarked the method on the ground-state charmonia and their first radial excitations, as well as on the
$\Omega_{ccc}$ baryon, all of which converge to the expected values. 
Applying the LQPM to the fully charmed tetraquark $cc\bar{c}\bar{c}$, we find that there existing $J^P=0^+$, $1^+$, $2^+$ fully charmed tetraquarks with masses 
$5980.7(14)~\mathrm{MeV}$, $6097.6(14)~\mathrm{MeV}$, 
and $6225.5(19)~\mathrm{MeV}$, respectively. They are about 
$13(1)~\mathrm{MeV}$,  $16(1)~\mathrm{MeV}$, and $30(2)~\mathrm{MeV}$ above their lowest strong decay thresholds, i.e. the $\eta_c\eta_c$, $\eta_cJ/\psi$, and $J/\psi J/\psi$ thresholds, respectively.
%unambiguously identified as a bound state. Since it lies below the lowest strong-decay channel, it can decay only through electromagnetic and weak interactions. 

LQPM has great potential for extension to larger multiquark systems, including pentaquark states. Such studies could provide forward-looking predictions for future experimental searches for exotic hadrons. In addition, the framework can accommodate potential models with a variety of interactions, including three-body forces, spin-orbit and tensor couplings, and other interactions beyond the simple two-body OGE plus confinement form, which are difficult to treat in conventional few-body approaches. It therefore offers a natural framework for investigating lattice-inspired potentials and exploring the confinement mechanism in tetraquark states, which could provide valuable insights into the internal structure of multiquark systems and into non-perturbative QCD.

{\it Acknowledgments}~~
We are grateful to Li-Sheng Geng, Lu Meng, Ya-Wen Pan, Jean-Marc Richard, Alfredo Valcarce, Javier Vijande, Shi-Lin Zhu, and Xian-Hui Zhong for valuable discussions. This work is supported, in part, by the National Natural Science Foundation of China with Grants Nos.~12342502, 12275259, 12375073,  12547105, and 12235018, the Science Challenge Project (No. TZ2025012),  and NSAF
No. U2330401.

%Bibliography
\nocite{*}
\bibliography{ref.bib}  

\onecolumngrid
\newpage
\appendix

\clearpage

\input{supp}

\end{document}

%% file: supp.tex
\section*{Supplemental Materials}
This supplement presents the implementation details of the finite volume effects, the original numerical results, the comparison with NNQS results under the same Hamiltonian, angular momentum and projection and the dependence of results on the lattice spacing.
%the dependence of results on the lattice spacing,
\section*{Finite-volume effects}
The matrix exact diagonalization scheme uses the Implicitly Restarted Lanczos Method~\cite{lehoucq:1998} to find the eigenvalues and eigenvectors by SciPy~\cite{Virtanen:2019joe} in Python. As this method becomes expensive for large volumes, we calculate the masses of the ground and the first radial excitations charmonia in Table~\ref{tab:meason_results}, using box sizes ranging from $L^3=1.5^3$ to 3.9$^3$ fm$^3$ and extrapolate the results to $L\to\infty$ by the energy shift function~\cite{Zhang:2025juc} in the finite volume,
\begin{equation}
    E_L = E_{\infty} + \frac{A_1}{L}e^{-A_2L}.
    \label{Eq:charmoniums_FVE}
\end{equation}
Here, $E_L$ is the results of charmonia in a box of size $L^3$ calculated by LQPM, $E_{\infty}$ is the masses in the infinite-volume limit, and $A_{1,2}$ are free parameters, with $A_{1}$ characterizing to the wave-function normalization coefficient, and $A_{2}$ scaling with the binding momentum. For the $\Omega_{ccc}$ case, our calculation results are in Table~\ref{tab:meason_results}, we use box sizes ranging from $L^3=1.5^3$ to 2.4$^3$ fm$^3$ and extrapolate the results to $L\to\infty$ by the energy shift function~\cite{Zhang:2026bqa} in the finite volume,
\begin{equation}
    E_L = E_{\infty} + \frac{A_1}{L^{3/2}}e^{-A_2L}.
    \label{Eq:barton_FVE}
\end{equation}
%The definition of the parameters is consistent with Eq.~\eqref{Eq:charmoniums_FVE}. 
For the calculation of fully charmed tetraquark $cc\bar{c}\bar{c}$, we also use Eq.~\eqref{Eq:barton_FVE} to extrapolate the lattice results for volumes larger than 1 fm.
Then, using non-linear least squares to fit the cost function $\chi^2=\sum_i(f(x_i)-y_i)^2$, where $f(x_i)$ is $E_L$ in Eq.~\eqref{Eq:charmoniums_FVE} or Eq.~\eqref{Eq:barton_FVE}, and $y_i$ are the results in Table~\ref{tab:meason_results} and Table~\ref{tab:cccc_1000_results}. All fitting results are shown in the Table~\ref{tab:fit_meason} and~\ref{tab:fit_tetra}. $A_1$, $A_2$ are strongly correlated. It should be noted that for charmonia and $\Omega_{ccc}$ data, using the Lüscher formula~\cite{Luscher:1985dn} for two-body systems and the extrapolation formula in the unitary limit~\cite{Meissner:2014dea} for three-body systems does not affect $E_{\infty}$. Specifically, by performing finite volume extrapolation using the Lüscher formula~\cite{Luscher:1985dn} and three-body system extrapolation formula~\cite{Meissner:2014dea}, the extrapolated results are $m_{\eta_c}=2984.3(4)~\mathrm{MeV}$, $m_{J/\psi}=3098.0(3)~\mathrm{MeV}$, $m_{\eta_c(2S)}=3634.5(2)~\mathrm{MeV}$, $m_{\psi(2S)}=3678.6(2)~\mathrm{MeV}$, and $m_{\Omega_{ccc}}=4816.7(13)~\mathrm{MeV}$.
\begin{table}[htbp]
\setlength{\tabcolsep}{8mm}
\renewcommand{\arraystretch}{2}
\centering
\caption{The fitting results of charmonia, $\Omega_{ccc}$.}
\begin{tabular}{ccccc}   
\hline
\hline
State&$E_{\infty}$ (MeV)&$A_1$ (MeV$^{0}$/MeV$^{-1/2}$)&$A_2$ (MeV) &$\chi^2$\\
\hline
$\eta_c$                  &2984.0(2) & $-$121.0(717) &873.8(768) &0.1\\
$J/\psi$                  &3097.8(2) & $-$93.2(284)  &766.5(394) &1.0\\
$\eta_c(2S)$              &3634.9(5) & $-$720.7(5246)&618.0(578) &0.6\\
$\psi(2S)$                &3679.4(5) & $-$460.9(2399) &562.7(414)&1.3\\
$\Omega_{ccc}$            &4817.3(9) & $-$45.2(282)   &852.9(743) &0.1\\
\hline
\hline
\end{tabular}
\label{tab:fit_meason}
\end{table}

% For the fitting results of the tetraquark state in Table~\ref{tab:fit_tetra}, we find that the finite volume extrapolation error grows quickly on data with smaller volumes, for example, the data with a lattice spacing $a=1/1400$ MeV$^{-1}$. 
%In the end, we chose the results with a lattice spacing $a=1/1000$ MeV$^{-1}$ as the finite volume extrapolation results.
%Taking it a step further, we use Eq.~\eqref{Eq:barton_FVE} to fit all the data at different lattice spacings with volumes larger than 1 fm (9 data points) together.  And the fitting results are shown in the Table~\ref{tab:fit_all_tetra}. Since the data we used for fitting includes results with different lattice spacings, the error comes not only from finite volume effects but also from the lattice spacing approaching the continuous limit. The uncertainties of Table~\ref{tab:fit_all_tetra} are determined by scanning over one parameter and minimizing the function with respect to all other parameters for each scan point using \texttt{mnprofile function} in iminuit.
\begin{table}[htbp]
\setlength{\tabcolsep}{8mm}
\renewcommand{\arraystretch}{2}
\centering
\caption{The fitting results of $cc\bar{c}\bar{c}$ in lattice simulation with lattice spacing $a=1/1000$ MeV$^{-1}$. }
%Since the number of data points with volumes greater than 1 fm is equal to the number of fitting parameters when the lattice spacing $a=1/1200$ and 1/1400 MeV$^{-1}$, the $\chi^2/d.o.f.$ is meaningless in this case.}
\begin{tabular}{ccccc}   
\hline
\hline
State&$E_{\infty}$ (MeV)&$A_1$ (MeV$^{-1/2}$)&$A_2$ (MeV)&$\chi^2/d.o.f.$\\
\hline
% \multicolumn{5}{c}{$a=1/1000$ MeV$^{-1}$}\\
$cc\bar{c}\bar{c}$ $(0^+)$&5980.7(14)& $-$1.6(1)     &526.9(104) &13.4\\
$cc\bar{c}\bar{c}$ $(1^+)$&6097.6(14)& $-$1.7(1)     &525.4(99) &15.3\\
$cc\bar{c}\bar{c}$ $(2^+)$&6225.5(19)& $-$1.1(1)     &400.6(79)  &25.7\\
% \hline
% \multicolumn{5}{c}{$a=1/1200$ MeV$^{-1}$}\\
% $cc\bar{c}\bar{c}$ $(0^+)$&5952.1(47) & $-$4.9(14)     &717.2(542) &$\cdots$\\
% $cc\bar{c}\bar{c}$ $(1^+)$&6072.4(44) & $-$6.1(16)     &742.0(492) &$\cdots$\\
% $cc\bar{c}\bar{c}$ $(2^+)$&6199.8(62)& $-$3.2(6)     &581.0(384)   &$\cdots$\\
% \hline
% \multicolumn{5}{c}{$a=1/1400$ MeV$^{-1}$}\\
% $cc\bar{c}\bar{c}$ $(0^+)$&5975.4(108)& $-$1.1(2)     &458.5(390) &$\cdots$\\
% $cc\bar{c}\bar{c}$ $(1^+)$&6091.1(104)& $-$1.2(2)     &462.1(361) &$\cdots$\\
% $cc\bar{c}\bar{c}$ $(2^+)$&6218.5(152)& $-$0.8(1)     &349.1(340) &$\cdots$\\
\hline
\hline
\end{tabular}
\label{tab:fit_tetra}
\end{table}

% \begin{table}[htbp]
% \setlength{\tabcolsep}{8mm}
% \renewcommand{\arraystretch}{2}
% \centering
% \caption{The fitting results of $cc\bar{c}\bar{c}$ in lattice simulation with all different lattice spacings.}
% \begin{tabular}{ccccc}   
% \hline
% \hline
% State&$E_{\infty}$ (MeV)&$A_1$ (MeV$^{-1/2}$)&$A_2$ (MeV)&$\chi^2/d.o.f.$\\
% \hline
% $cc\bar{c}\bar{c}$ $(0^+)$&5929.4$^{+16.6}_{-12.5}$& $-$5.0$^{+15.7}_{-3.5}$     &752.0$^{+256.5}_{-226.5}$ &\\
% $cc\bar{c}\bar{c}$ $(1^+)$&6066.7$^{+12.0}_{-10.0}$& $-$3.1$^{+3.4}_{-1.6}$     &642.5$^{+134.5}_{-126.0}$ &\\
% $cc\bar{c}\bar{c}$ $(2^+)$&6215.6$^{+15.6}_{-12.9}$& $-$1.4$^{+0.7}_{-0.4}$     &435.5$^{+76.5}_{-75.0}$ &\\
% \hline
% \hline
% \end{tabular}
% \label{tab:fit_all_tetra}
% \end{table}

\section*{Lattice results}

The lattice method is first benchmarked on the charmonia and $\Omega_{ccc}$. The calculations are performed at lattice spacing $a=1/2000$ MeV$^{-1}\approx0.1$ fm with the meson box sizes ranging from $L^3=1.5^3$ to 3.9$^3$ fm$^3$, and the baryon box sizes from $L^3=1.5^3$ to 2.4$^3$ fm$^3$. The results are shown in the Table~\ref{tab:meason_results}. Afterwards, we extend our method to fully charmed tetraquark $cc\bar c\bar c$. We performed in lattice spacing $a=1/1000$ MeV$^{-1}\approx0.2$ fm with the box sizes ranging from $L^3=1.0^3$ to 1.8$^3$ fm$^3$, and the results are shown in Table~\ref{tab:cccc_1000_results}.
\begin{table}[htbp]
\setlength{\tabcolsep}{8mm}
\renewcommand{\arraystretch}{2}
\centering
\caption{Predicted mass (MeV) spectra of the charmonia and
all-charmed baryons $\Omega_{ccc}$.}
\begin{tabular}{cccccc}   
\hline
\hline
$L/a$& $\eta_c$&$J/\psi$&$\eta_c(2S)$&$\psi(2S)$&$\Omega_{ccc}$\\
\hline

15   & 2961.1  & 3058.5  & 3494.9  & 3557.9  & 4719.2  \\
16   & 2970.0  & 3072.3  & 3501.7  & 3556.2  & 4754.0  \\
17   & 2975.3  & 3081.1  & 3508.1  & 3555.7  & 4776.4  \\
18   & 2978.8  & 3087.4  & 3523.3  & 3566.2  & 4792.8  \\
19   & 2980.9  & 3091.2  & 3535.6  & 3574.9  & 4802.3  \\
20   & 2982.2  & 3093.8  & 3552.2  & 3589.6  & 4808.6  \\
21   & 2982.9  & 3095.3  & 3565.2  & 3601.2  & 4812.0  \\
22   & 2983.3  & 3096.3  & 3579.8  & 3615.6  & 4814.1  \\
23   & 2983.6  & 3096.9  & 3590.8  & 3626.6  & 4815.2  \\
24   & 2983.7  & 3097.2  & 3601.7  & 3638.5  & 4815.9  \\
25   & 2983.8  & 3097.4  & 3609.7  & 3647.3  & $\cdots$ \\
26   & 2983.9  & 3097.6  & 3616.8  & 3655.7  & $\cdots$  \\
27   & 2983.9  & 3097.6  & 3621.8  & 3661.6  & $\cdots$  \\
28   & 2983.9  & 3097.7  & 3625.9  & 3666.8  & $\cdots$  \\
29   & 2983.9  & 3097.7  & 3628.6  & 3670.3  & $\cdots$  \\
30   & 2983.9  & 3097.7  & 3630.6  & 3673.1  & $\cdots$  \\
31   & 2983.9  & 3097.7  & 3631.9  & 3674.9  & $\cdots$  \\
32   & 2983.9  & 3097.7  & 3632.9  & 3676.3  & $\cdots$  \\
33   & 2983.9  & 3097.7  & 3633.5  & 3677.1  & $\cdots$  \\
34   & 2983.9  & 3097.7  & 3633.9  & 3677.7  & $\cdots$  \\
35   & 2983.9  & 3097.7  & 3634.1  & 3678.1  & $\cdots$  \\
36   & 2983.9  & 3097.7  & 3634.3  & 3678.3  & $\cdots$  \\
37   & 2983.9  & 3097.7  & 3634.4  & 3678.5  & $\cdots$  \\
38   & 2983.9  & 3097.7  & 3634.4  & 3678.6  & $\cdots$  \\
39   & 2983.9  & 3097.7  & 3634.5  & 3678.6  & $\cdots$  \\
\hline
\hline
\end{tabular}
\label{tab:meason_results}
\end{table}

\begin{table}[htbp]
\setlength{\tabcolsep}{6mm}
\renewcommand{\arraystretch}{2}
\centering
\caption{Predicted mass (MeV) spectra of fully charmed tetraquark $cc\bar{c}\bar{c}$ in a lattice simulation with lattice spacing $a=1/1000$ MeV$^{-1}$.}
\begin{tabular}{ccccccc}   
\hline
\hline
$cc\bar{c}\bar{c}$&$L/a$& 5& 6& 7& 8&9\\
\hline 
\multirow{3}{*}{$J^P=0^+$}& Mass& 5658.4  & 5836.5  & 5910.3 & 5951.9 & 5962.5\\
& $\chi_{6\otimes\bar{6}}$& $\sim57.4\%$& $\sim60.7\%$& $\sim61.9\%$& $\sim63.9\%$ & $\sim65.1\%$\\
& $\chi_{\bar{3}\otimes3}$& $\sim42.6\%$& $\sim39.3\%$& $\sim38.1\%$& $\sim36.1\%$ & $\sim34.9\%$ \\
\multirow{3}{*}{$J^P=1^+$}& Mass& 5756.9  & 5945.2  & 6022.8 & 6067.0 & 6078.4\\
&
$\chi_{6\otimes\bar{6}}$&$\sim12.7\%$&$\sim28.0\%$&$\sim44.0\%$&$\sim56.3\%$&$\sim61.6\%$\\
& $\chi_{\bar{3}\otimes3}$&$\sim87.3\%$&$\sim72.0\%$&$\sim56.0\%$&$\sim43.7\%$&$\sim38.4\%$\\
\multirow{3}{*}{$J^P=2^+$}& Mass& 5811.4  & 6014.5  & 6110.3 & 6169.7 & 6187.9\\
& $\chi_{6\otimes\bar{6}}$&$\sim8.1\%$&$\sim19.6\%$&$\sim34.6\%$&$\sim50.5\%$&$\sim58.7\%$\\
& $\chi_{\bar{3}\otimes3}$&$\sim91.9\%$&$\sim80.4\%$&$\sim65.4\%$&$\sim49.5\%$&$\sim41.3\%$\\
\hline
\hline
\end{tabular}
\label{tab:cccc_1000_results}
\end{table}

\begin{figure*}[htbp]
    \centering
    \includegraphics[width=0.44\textwidth]{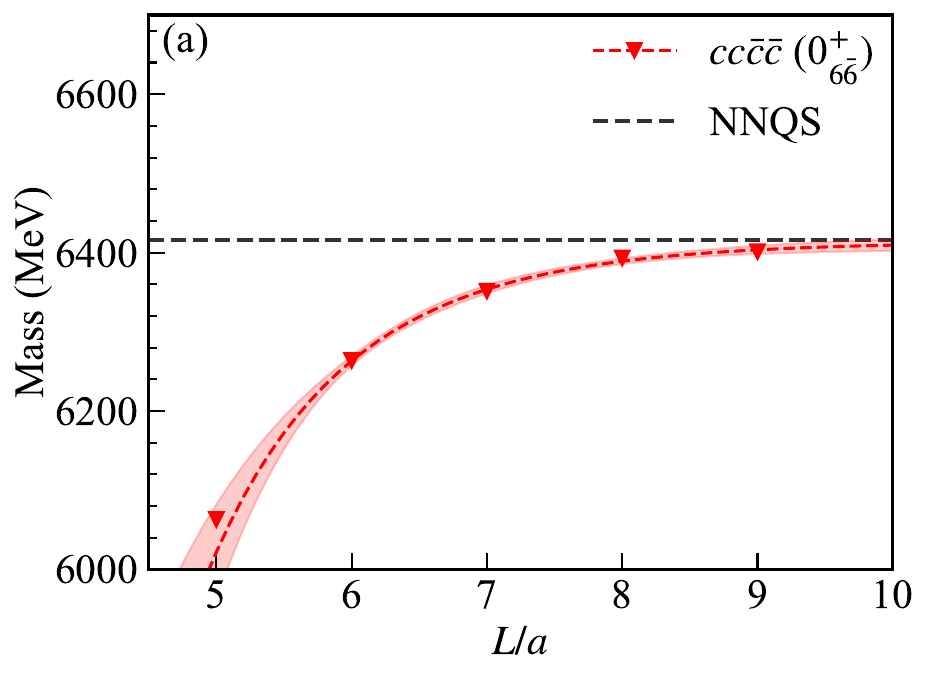}
    \includegraphics[width=0.44\textwidth]{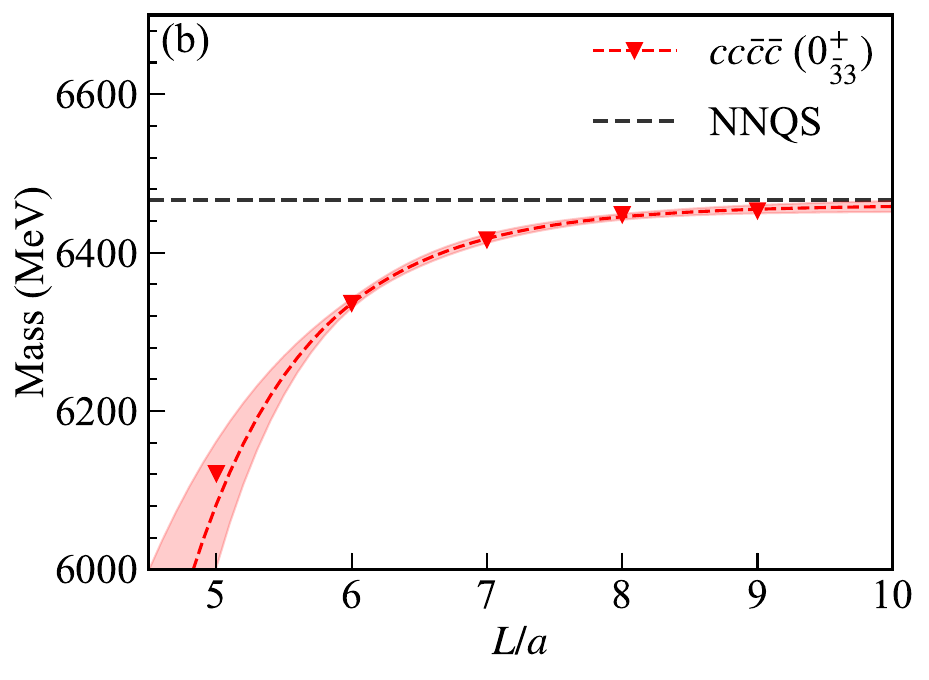}
    \includegraphics[width=0.44\textwidth]{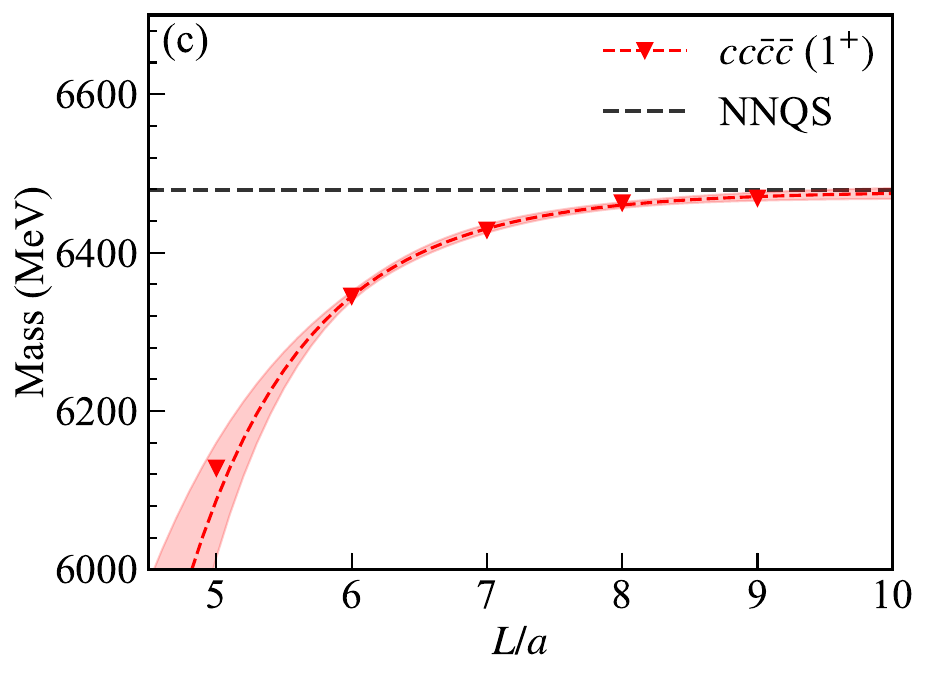}
    \includegraphics[width=0.44\textwidth]{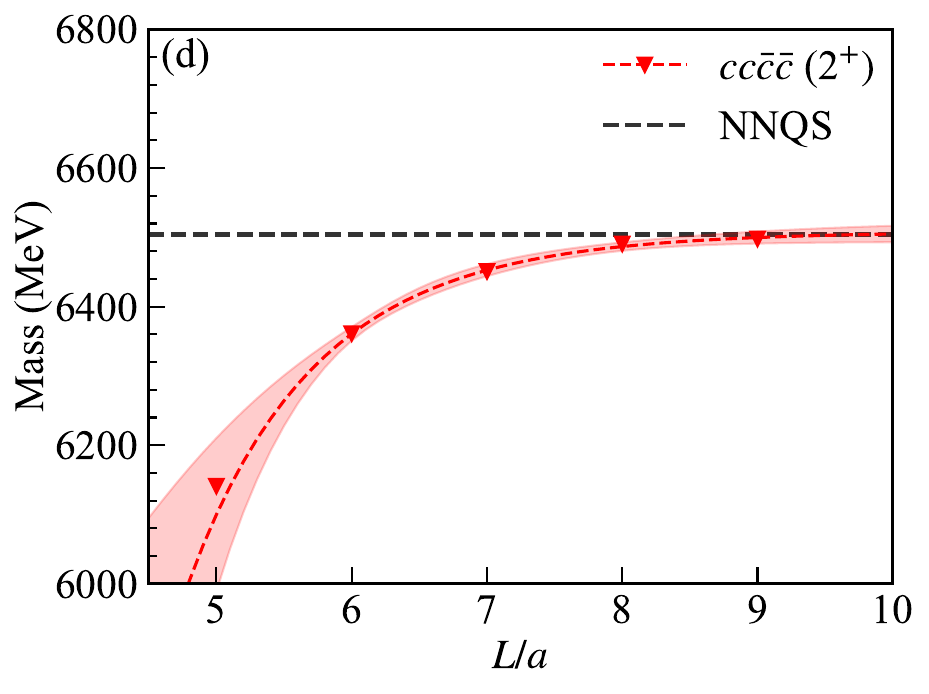}
    \caption{The predicted masses of S-wave fully charmed tetraquark ground states with $J^P=0^+_{6\bar{6}}$(a), $0^+_{\bar{3}3}$ (b), $1^+$ (c), and $2^+$ (d) in a lattice simulation with lattice spacing $a=1/800$ MeV$^{-1}$ in a comparison with the NNQS results~\cite{Li:2026hol}. The red triangles represent this work predictions. The finite-volume errors are shown by the red shaded area, black dashed lines in subfigures represent the NNQS results~\cite{Li:2026hol}. }
    \label{fig:tetraquark_test}
\end{figure*}

\section*{Comparison with NNQS results under the same Hamiltonian}
Under the same Hamiltonian, the huge difference between our results and the NNQS results~\cite{Li:2026hol} is because they only construct antisymmetric color-spin wavefunctions and symmetric spatial wavefunctions, while we enforce antisymmetry by explicitly summing over all possible cases. The contribution of higher partial waves has affected the ground state energy. In order to illustrate this issue, we also performed calculations considering only the case where the spatial wave function is symmetric. Specifically, we performed in a lattice spacing $a=1/800$ MeV$^{-1}\approx0.25$ fm with the box sizes ranging from $L^3=1.25^3$ to 2.25$^3$ fm$^3$, and results are extrapolated using the finite-volume effect formula. The results are shown in Fig.~\ref{fig:tetraquark_test}. Table~\ref{tab:tetraquark_test} shows a comparison of our LQPM results with the NNQS results, and they are consistent within their uncertainties.
\begin{table}[htbp]
\setlength{\tabcolsep}{9mm}
\renewcommand{\arraystretch}{2}
\centering
\caption{Predicted mass (MeV) spectra of fully charmed tetraquark $cc\bar{c}\bar{c}$ in a lattice simulation with lattice spacing $a=1/800$ MeV$^{-1}$. Our results are compared with
NNQS results.}
\begin{tabular}{ccccc}   
\hline
\hline
$cc\bar{c}\bar{c}$&$J^P=0^+_{6\bar{6}}$&$J^P=0^+_{\bar{3}3}$&$J^P=1^+$&$J^P=2^+$\\
\hline

This work& 6414.0(35)& 6460.2(28)& 6477.1(29)& 6508.3(32)\\
NNQS     & 6416.4(8)& 6466.2(7)& 6479.0(7)& 6503.8(8)\\
\hline
\hline
\end{tabular}
\label{tab:tetraquark_test}
\end{table}

\section*{Angular momentum and parity projection}

To probe the quantum numbers of the states of interest, 
the standard angular momentum and parity projection (AMP) technique is applied to the calculated eigenstate $|\Psi_0\rangle$. 
The essential tool in constructing the irreps is the angular momentum projection formula~\cite{Lu:2014xfa}
\begin{align}
    |\Psi_A\rangle=&\frac{d_n}{24}\sum_{i=1}^{24}\chi_n(\Omega_i)R(\Omega_i)|\Psi_0\rangle,
\end{align}
where $|\Psi_0\rangle$ is a given eigenstate. Here $n=1,2,3,4,5$ corresponds to the five irreducible representations of the cubic rotational group, i.e. $A_1$, $A_2$, $E$,  $T_1$, and $T_2$. $d_n$ is the dimension of the corresponding irreducible representation. Their relation to the continuum SO(3) rotational group is presented in Tab.~\ref{tab:irreps}. $R(\Omega_i)$ is the $i$th rotational operator with $\Omega_i$ (with $i=1,2,\cdots 24$) the $i$th cubic rotational element, produced by the products of the $\pi/2$ rotation along the $x,y,z$ axes. $\chi_n$ is the character of the $n$th irreducible representation. We apply the AMP to the ground and first excited states and find that all of them are $S$-wave contributions, i.e. $A_1$. 
\begin{table}[htbp]
\setlength{\tabcolsep}{9mm}
\renewcommand{\arraystretch}{2}
\centering
\caption{Continuum total angular momenta decomposed into $irreps$(dim) of the $O$ group.}
\begin{tabular}{cc}
    \hline
    \hline
     SO(3)&$O$\\
    \hline
     J=0& $A_1(1)$\\
     J=1& $T_1(3)$\\
     J=2& $E(2)\oplus T_2(3)$\\
     J=3& $T_1(3)\oplus T_2(3)\oplus A_2(1)$\\
     J=4& $A_1(1)\oplus T_1(3)\oplus E(2)\oplus T_2(3)$\\
    \hline
    \hline
    \end{tabular}  \label{tab:irreps}
\end{table}

\section*{Dependence on the lattice spacing}
\begin{figure*}[htbp]
    \centering
    \includegraphics[width=0.44\textwidth]{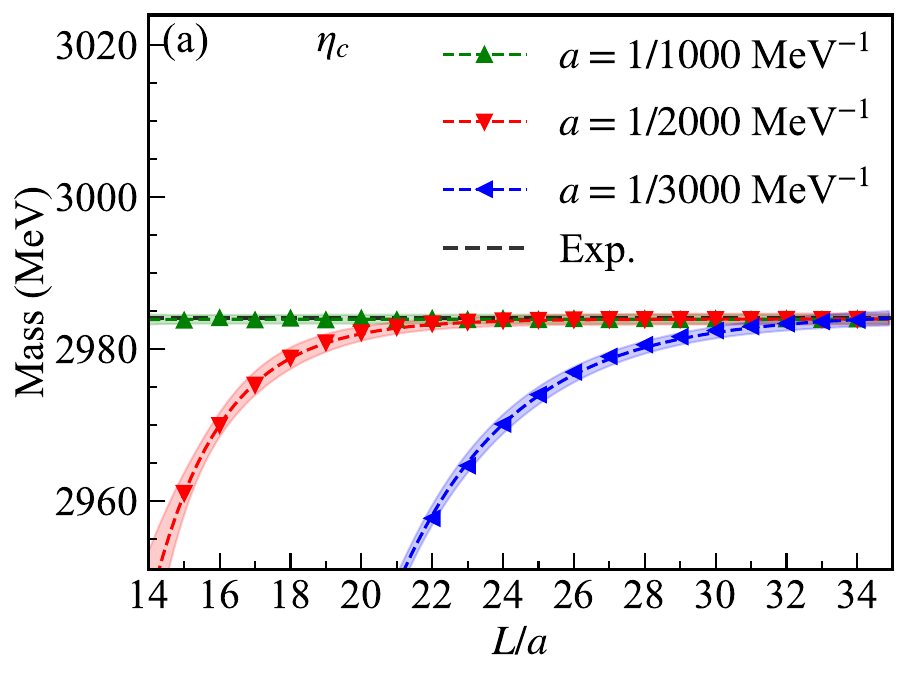}
    \includegraphics[width=0.44\textwidth]{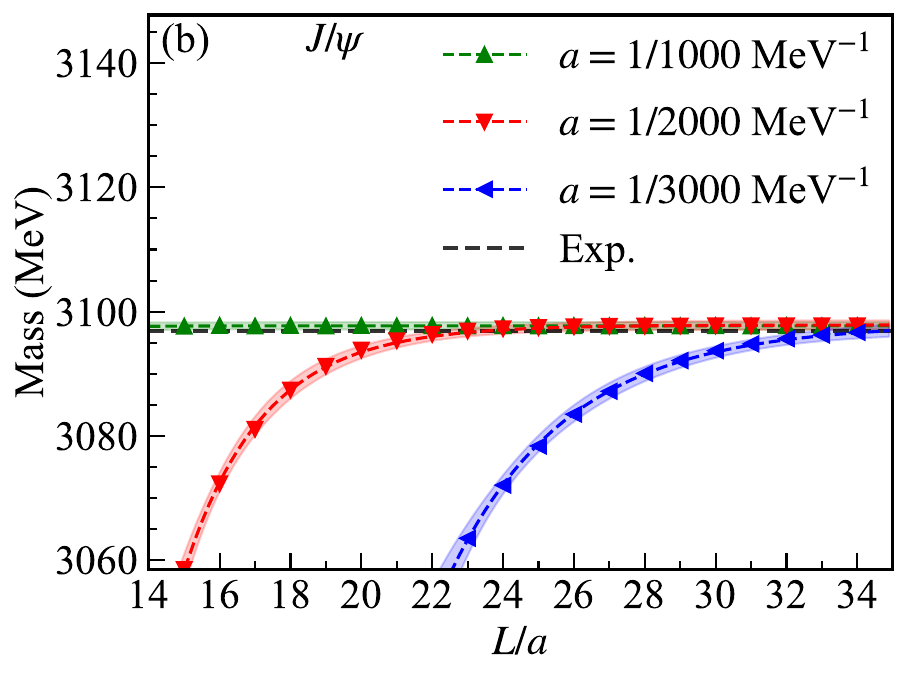}
    \includegraphics[width=0.44\textwidth]{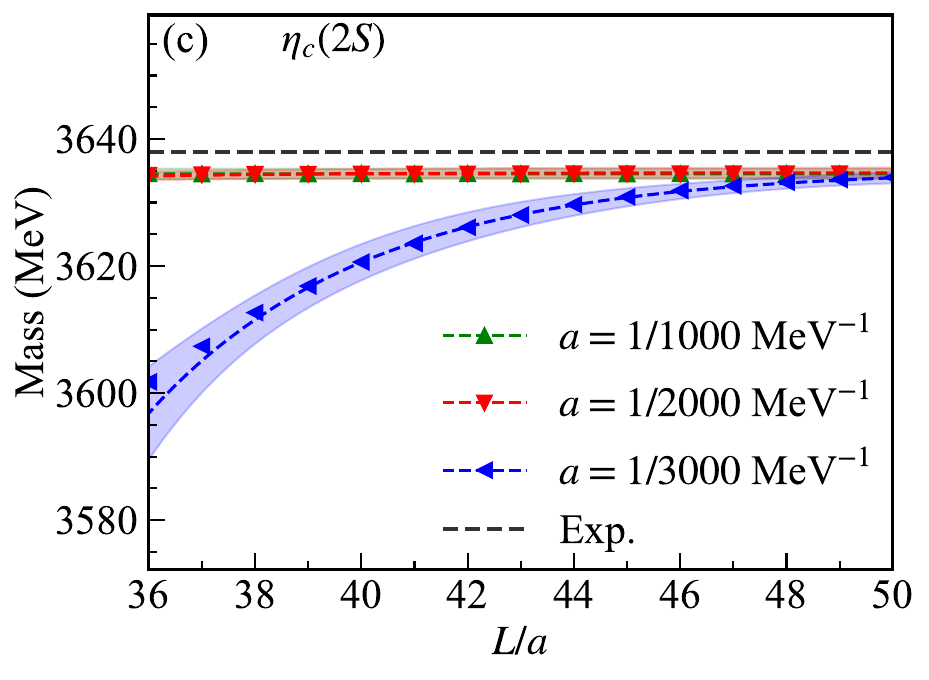}
    \includegraphics[width=0.44\textwidth]{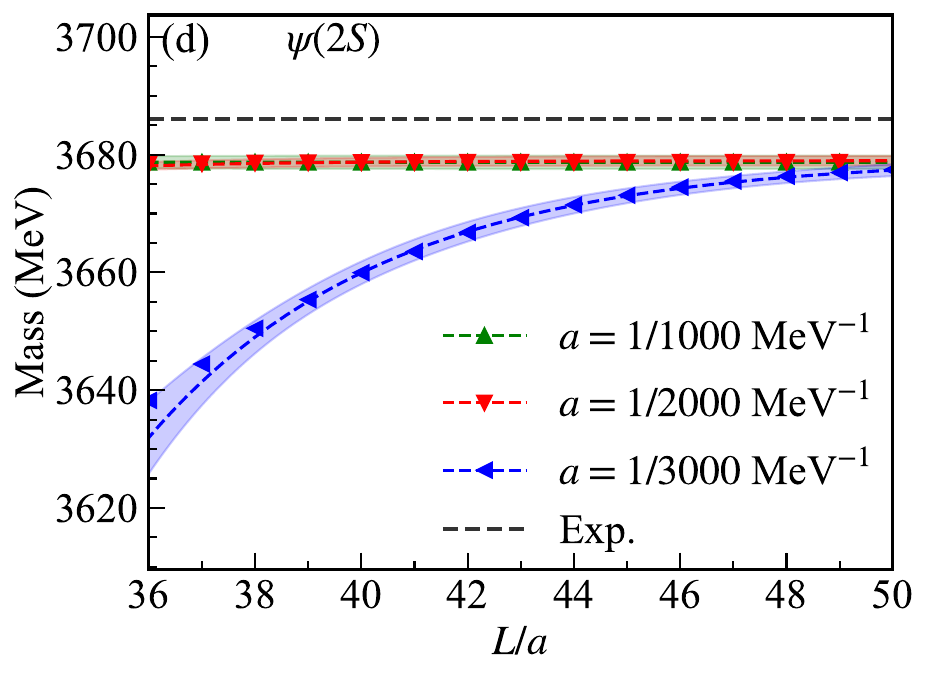}
    \caption{The predicted masses of $\eta_c$(a), $J/\psi$ (b), $\eta_c(2S)$ (c), and $\psi(2S)$ (d) in a lattice simulation with lattice spacing $a=1/1000$, 1/2000 and 1/3000 MeV$^{-1}$. The triangles represent this work predictions. The finite-volume errors are shown by the shaded area, black dashed lines represent experimental data. }
    \label{fig:meason_dep_a}
\end{figure*}

For charmonia, the lattice spacing is already small enough, it is not necessary to consider the dependence on the lattice spacing. We perform at lattice spacing $a=1/1000$, 1/2000 and 1/3000 MeV$^{-1}\approx0.2$, 0.1 and 0.07 fm. The results are shown in the Fig.~\ref{fig:meason_dep_a}. When $a=1/1000$ MeV$^{-1}$, to reduce the dependence of results on the lattice spacing, we refitted the parameters of interaction and results are shown in Table~\ref{tab:parameters_new}, the calculated results converge to the experimental data, and when $a=1/2000$ MeV$^{-1}$ or even smaller lattice spacing, we use the parameters in Table~\ref{tab:parameters}, the calculated results converge to the experimental data. This indicates that for charmonia, the lattice spacing $a=1/2000$ MeV$^{-1}$ is appropriate.
\begin{table}[htbp]
\setlength{\tabcolsep}{3.5mm}
\renewcommand{\arraystretch}{1.5}
\centering
\caption{Parameters of the model when lattice spacing $a=1/1000$ MeV$^{-1}$.}
\label{tab:parameters_new}
\begin{tabular}{cccc}
\hline
\hline
$m_c$ (GeV) & $\alpha_{cc}$ & $\sigma_{cc}$ (GeV) & $b$ (GeV$^2$)\\
\hline
 1.491      &  0.5508        & 1.0403           & 0.1413\\
\hline
\hline
\end{tabular}
\end{table}

\begin{figure*}[htbp]
    \centering
    \includegraphics[width=0.44\textwidth]{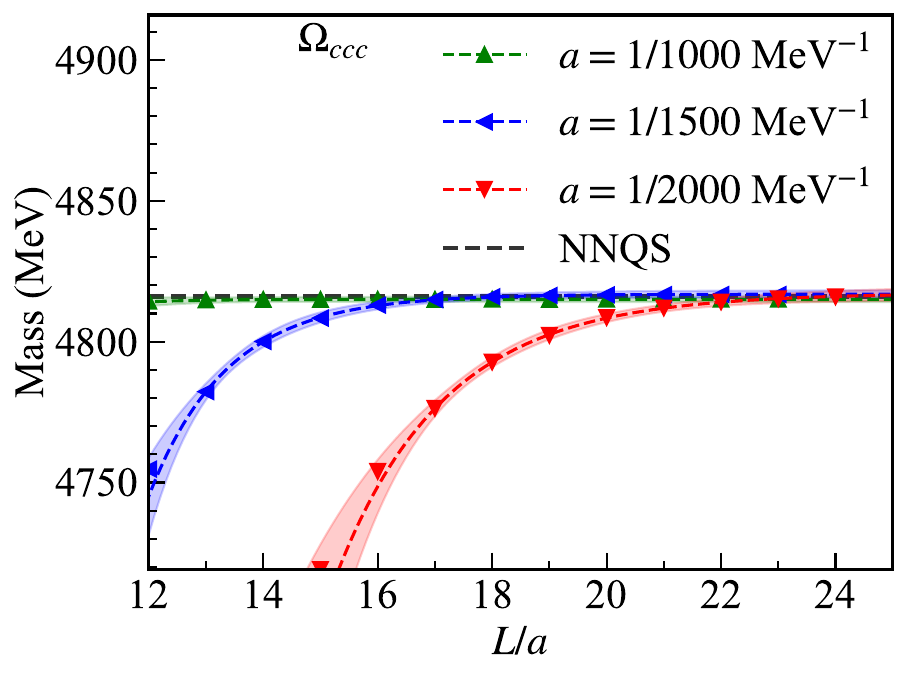}
    \caption{The predicted masses of $\Omega_{ccc}$ in a lattice simulation with lattice spacing $a=1/1000$, 1/1500 and 1/2000 MeV$^{-1}$. The triangles represent this work predictions. The finite-volume errors are shown by the shaded area, black dashed lines represent NNQS result. }
    \label{fig:baryon_dep_a}
\end{figure*}
For $\Omega_{ccc}$, we perform at lattice spacing $a=1/1000$, 1/1500 and 1/2000 MeV$^{-1}\approx0.2$, 0.13 and 0.1 fm. The results are shown in Fig.~\ref{fig:baryon_dep_a}. When $a=1/1000$ MeV$^{-1}$ or even smaller lattice spacing, the results converge to the NNQS result. This indicates that for the $\Omega_{ccc}$, $a=1/2000$ MeV$^{-1}$ is appropriate. For fully charmed tetraquark $cc\bar c\bar c$, due to the limitations of finite volume effects, direct calculations with lattice spacing smaller than $a=1/1000$ MeV$^{-1}$ does not converge. To overcome this problem, we refit the parameters listed in Table~\ref{tab:parameters} to the charmonium spectrum. This is the standard strategy in Nuclear Lattice EFT\cite{Niu:2025uxk}. The quark mass, for instance, can be understood as a scale-dependent running quark mass.

%Both the convergence of the $\Omega_{ccc}$ results at $a=1/1000$ MeV$^{-1}$ and the agreement with the NNQS results obtained under the same Hamiltonian suggest that the lattice spacing of $a=1/1000$ MeV$^{-1}$ is appropriate for the $cc\bar{c}\bar{c}$ calculation. 